\documentclass[aps, superscriptaddress,  twocolumn]{revtex4}
\usepackage{braket}
\usepackage{graphicx}
\usepackage{dcolumn}
\usepackage{bm}
\usepackage{svg}
\usepackage{amsmath}
\usepackage{booktabs}
\usepackage{array}
\usepackage{multirow}
\usepackage{float}
\usepackage{lipsum}
\usepackage{gensymb}

\begin{document}

\preprint{APS/123-QED}

 \title{Talbot effect-based sensor measuring grating period change in subwavelength range}

\author{Saumya J. Sarkar}
\email{saumya_js@iitgn.ac.in}
 \affiliation{Photonic Sciences Lab., Physical Research Laboratory, Navrangpura, Ahmedabad 380009, Gujarat, India}
\affiliation{Indian Institute of Technology-Gandhinagar, Ahmedabad 382424, Gujarat, India}

 \author{M. Ebrahim-Zadeh}
 \affiliation{ICFO-Institut de Ciencies Fotoniques, The Barcelona Institute of Science and Technology, 08860 Castelldefels (Barcelona), Spain}
\affiliation{Institucio Catalana de Recercai Estudis Avancats (ICREA), Passeig Lluis Companys 23, Barcelona 08010, Spain}

 \author{G. K. Samanta}
 \affiliation{Photonic Sciences Lab., Physical Research Laboratory, Navrangpura, Ahmedabad 380009, Gujarat, India}%

\date{\today}

\begin{abstract}
\noindent Talbot length, the distance between two consecutive self-image planes along the propagation axis for a periodic diffraction object (grating) illuminated by a plane wave, depends on the period of the object and the wavelength of illumination. This property makes the Talbot effect a straightforward technique for measuring the period of a periodic object (grating) by accurately determining the Talbot length for a given illumination wavelength. However, since the Talbot length scale is proportional to the square of the grating period, traditional Talbot techniques face challenges when dealing with smaller grating periods and minor changes in the grating period. Recently, we demonstrated a Fourier transform technique-based Talbot imaging method that allows for controlled Talbot lengths of a periodic object with a constant period and illumination wavelength. Using this method, we successfully measured periods as small as a few micrometers and detected sub-micrometer changes in the periodic object. Furthermore, by measuring the Talbot length of gratings with varying periods imaged through the combination of a thick lens of short focal length and a thin lens of long focal length and large aperture, we determined the effective focal length of the thick lens in close agreement with the theoretical effective focal length of a thick lens in the presence of spherical aberration. These findings establish the Talbot effect as an effective and simple technique for various sensing applications in optics and photonics through the measurement of any physical parameter influencing the Talbot length of a periodic object.
\end{abstract}




\maketitle


\noindent 
In 1836, H. F. Talbot discovered a remarkable self-imaging phenomenon while studying periodic structures. A spatially coherent wave illuminating a periodic object produces an image of the object, without any imaging system, at regular intervals along the propagation direction. The image planes are known as Talbot planes \cite{talbot1836lxxvi}, and the distance between two consecutive Talbot planes is called Talbot length. The Talbot length as derived by Lord Rayleigh \cite{rayleigh1881xxv} is represented as,

\begin{equation}\label{Rayleigh_eqn}
z_{T}=\frac{\lambda}{1-\sqrt{\left(1-\frac{\lambda^{2}}{\Lambda^{2}}\right)}} \approx \frac{2 \Lambda^{2}}{\lambda}.
\end{equation}

Here, $\Lambda$ is the grating period, and $\lambda$ is the wavelength of the incident radiation. At half of the Talbot length, the image has a lateral shift by an amount of $\Lambda$/2 \cite{Berry:96}.  

However, the birth of the laser in 1960 has revitalized interest in the Talbot self-imaging effect. The elegance and simplicity of the Talbot effect have inspired a surge of studies in coherent optical signal processing, leading to a wide range of innovative applications that not only enrich our fundamental understanding of diffraction and its related optical phenomena but also extend to various other fields of science \cite{wen2013talbot}. In recent times, the Talbot effect has been utilized to build measurement devices for various physical parameters, including spectrometry to measure the spectrum of an optical source in a compact footprint through the Fourier transform of Talbot lengths \cite{kung2001transform, ye2016miniature}, wavefront sensing \cite{Podanchuk:16} and aberration measurement \cite{Wang:09}, precise displacement, angle and tilt measurements \cite{spagnolo2000talbot,Cao:22,TESTORF1996167}. Wavefront aberrations have been detected using the Talbot effect through various methods \cite{sekine2006measurement, podanchuk2016adaptive}. Additionally, aberration-like features have been observed in the post-paraxial Talbot effect \cite{ring2012aberration}. Similarly, the X-ray Talbot interferometry has been employed to measure very small grating spacings and shapes for small-angle scattering \cite{sunday2015determining}. In the strong imaging regime, where frequencies exceed $1/\lambda$, superoscillation enables subwavelength imaging with grating waves \cite{Deng:20}, offering significant advantages over traditional imaging techniques. Over the past decade, numerous efforts have been made to develop super-oscillatory lenses and harness their capabilities for super-resolution or subwavelength imaging \cite{rogers2012super, huang2007focusing}. Since the Talbot length, as shown by Eq. \ref{Rayleigh_eqn}, is a function of the grating period, it can, in principle, be used as a measuring quantity to determine the unknown period of an object. In fact, efforts have been made to achieve high-precision grating period measurement using two-grating Talbot interferometers, successfully reaching a lower bound of 30 µm for the measured grating period \cite{Photia:19}. However, when detecting an object period comparable to the wavelength, the non-paraxial Talbot effect requires the use of a more complex confocal microscope setup \cite{ARRIETA2019988}, thereby diminishing the inherent simplicity of the Talbot effect. It is also important to note that the approximation used in calculating the Talbot length, as given by Eq.~\ref{Rayleigh_eqn}, does not hold in the non-paraxial Talbot effect, where the grating period is comparable to the incident wavelength, \cite{Berry:96, hua2012talbot}. Therefore, it is imperative to develop new experimental schemes that can detect subwavelength variations in periodicity through the direct measurement of Talbot lengths, all while preserving the simplicity of the experimental setup. 

Here, we report on an experimental scheme, supported by a robust theoretical framework, to produce tunable Talbot lengths even for objects with small grating periods using standard optical components available in any optics lab and establish a Talbot-based sensor capable of measuring both small absolute grating periods and changes in periodicity within the subwavelength range. Additionally, we utilized the non-paraxial Talbot effect, observed for a grating period comparable to the wavelength of the illuminating light, to detect spherical aberration in a thick lens. This approach enabled us to measure changes in the focal length of the thick lens of small focal length under non-paraxial illumination, both theoretically and experimentally.


 
\begin{figure}[h]
\centering
\includegraphics[width=\linewidth]{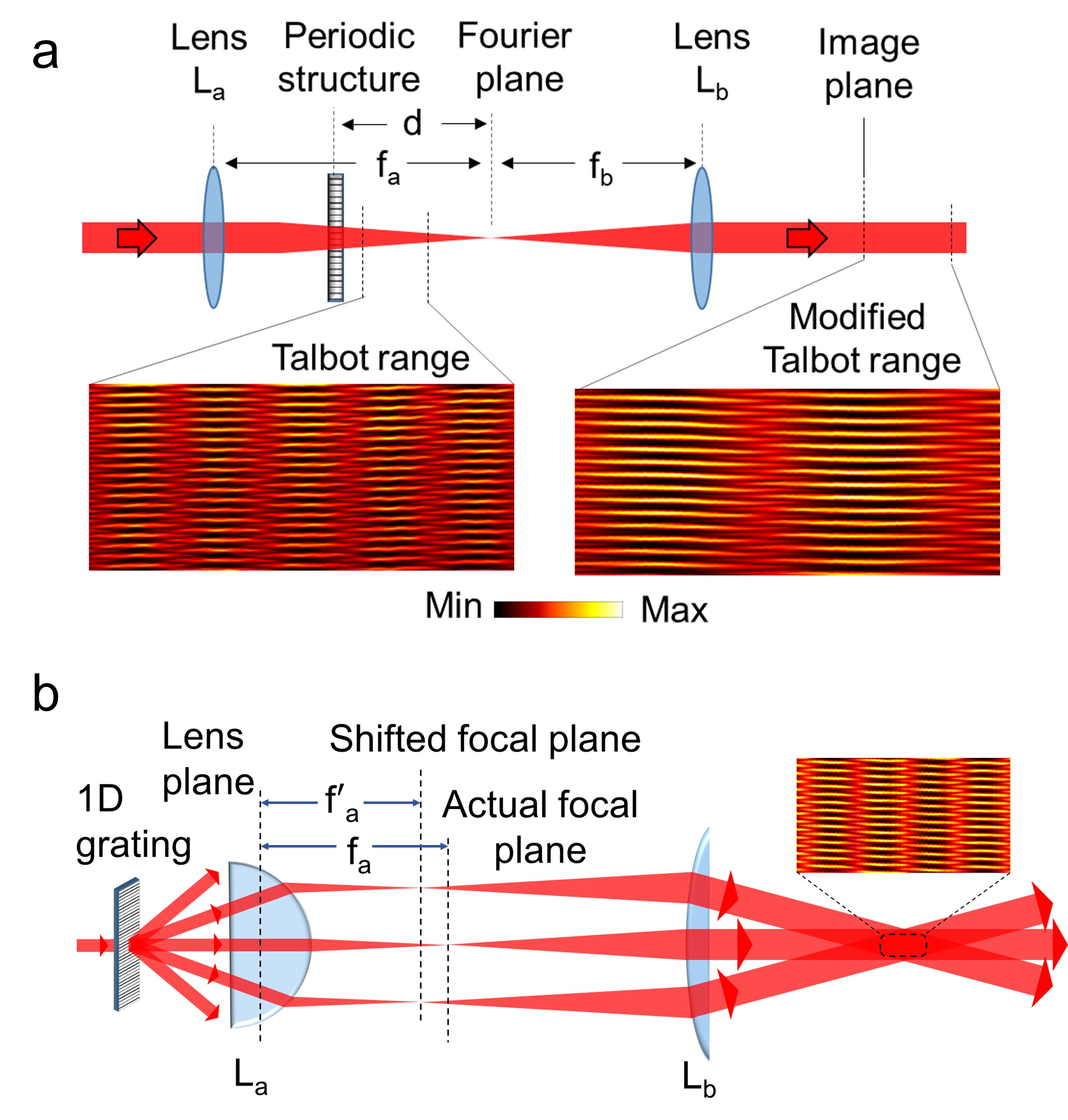}
\caption{Schematic representation of the experimental setup for a. verification of tunable Talbot length and measurement of small change in the grating period of the object, and b. measurement of the effective focal length of thick less using the non-paraxial Talbot effect. $L_a$ and $L_b$ are lenses of focal lengths, $f_a$ and $f_b$. $d$ is the position of the object away from the back focal plane of the lens, $L_a$. The objects are gratings. Inset: light carpet measured in the experiments.}
\label{Figure 1}
\end{figure}

The concepts of the tunable Talbot length for sensing small variations in periodicity and the measurement of the effective focal length change or the spherical aberration of a thick lens are pictorially shown in Fig. \ref{Figure 1}. In both cases, we have used a combination of lens $L_a$ of focal lengths, $f_a$ for Fourier transformation and $L_b$ of focal lengths, $f_a$, for inverse Fourier transformation to image \cite{goodman2005introduction} the object for the measurement of the Talbot images along propagation, avoiding the mechanical constraints of the experiment. However, the object's position changes depending on the need. To explore the tunable Talbot length as reported previously \cite{Harshit24}, the periodic object is placed after the lens $L_a$, at a distance $d$ from its back focal plane (Fourier plane), whereas, for the measurement of the effective focal length of the thick lens, the object is placed at the front focal plane of the lens, $L_a$. 

%
According to Fourier transform theory \cite{goodman2005introduction}, adjusting the position (distance, $d$) of the periodic object from the back focal plane of the lens, $L_a$, we can control the distribution of the spatial frequency of the object at the Fourier plane of lens $L_{a}$ due to the variation in the illuminated region of the object. In fact, it has been observed \cite{Harshith2019} that any periodic object results in an array of spots at the Fourier plane, and the spatial distribution of the array depends on the position of the periodic object away from the back focal plane of the lens. Each spot of the array has an intensity profile the same as the input beam (here, Gaussian beam). The modified pitch of the Gaussian beam array at the Fourier plane of the lens, $L_{a}$, due to the period object of pitch $\Lambda$ placed at a distance $d$ from the back focal plane can be represented as \cite{Jabir2016},
\begin{equation}\label{Eq. 2}
\Lambda^{\prime} =\frac{\lambda d }{\Lambda}.
\end{equation}
Here, $\lambda$ is the wavelength of the beam illuminating the object. The inverse Fourier transform of the Gaussian beam array using the lens $L_b$ results in the image of the period object at the image plane with effective pitch or period as
\begin{equation}\label{Eq. 3}
\Lambda_{eff} =\frac{f_b \lambda}{\Lambda^{\prime}} = \frac{f_b}{d} \Lambda.
\end{equation}
As a result, the Talbot length, $Z_T = 2\Lambda^2/\lambda$, of the object of the period, $\Lambda$, measured right after the object (see the inset Talbot range of Fig. \ref{Figure 1}) will be modified for the measured Talbot length after the image plane (see the inset Modified Talbot range of Fig. \ref{Figure 1}). Using Eq. \ref{Eq. 2} and \ref{Eq. 3} we can find the modified Talbot length as,
\begin{equation}\label{Eq. 4}
Z_T^M = \frac{2\Lambda_{eff}^2}{\lambda} = \left(\frac{f_b}{d}\right)^2 \left(\frac{2\Lambda^2}{\lambda}\right)= \left(\frac{f_b}{d}\right)^2 Z_T.  
\end{equation}
Using Eq.~\ref{Eq. 4}, we can derive the change in modified Talbot length $\Delta Z_T^M$ due to the change of grating period ($\Delta \Lambda$) as 
\begin{equation}\label{Eq. 5}
\Delta Z_T^M = \left(\frac{4\Lambda}{\lambda}\right) \left(\frac{f_b}{d}\right)^2 \Delta \Lambda.
\end{equation}
It is evident from Eq.~\ref{Eq. 4} and ~\ref{Eq. 5}, that the modified Talbot length, $Z_{T}^M$ and its change, $\Delta Z_T^M$, with change in the periodicity, ($\Delta \Lambda$), are modulated with a factor $(f_b/d)^{2}$, square of the ratio of the focal length of the inverse Fourier transforming lens, $L_{b}$, and the position of the periodic object away from the back focal plane of the Fourier transforming lens, $L_{a}$. Therefore, for a fixed periodicity, ($\Lambda$), of the object, focal length, ($f_b$),  of the lens, and illuminating wavelength, ($\lambda$), one can access variable Talbot length and determine the small variation of the grating period by simply adjusting the position of the object (i.e., the value of $d$). 
Again, from Eq. ~\ref{Eq. 5}, it is evident that, for a fixed value of $f_b/d$, wavelength, $\lambda$, and change in Talbot length, $\Delta Z_T^M$, the $\Delta \Lambda$ is inversely proportional the period, $\Lambda$, of the object. Therefore, one can achieve higher resolution (smaller $\Delta \Lambda$) for longer grating periods, $\Lambda$. On the other hand, one has to select the value of $d$, such that the spot size of the beam illuminating the periodic object should be larger than the period of the object to ensure the Talbot effect. Given these limitations, we had to limit the period of the grating to a certain level. 

However, using the objects of lower grating periods, we have studied the change of effective focal length of the thick lens due to the spherical aberration under non-paraxial illumination.  
Under paraxial approximation, a lens yields a multiplicative phase transformation factor given by $U(x, y)\sim \exp \left[\frac{i k}{2 f_{a}}\left(x^{2}+y^{2}\right)\right]$, where x and y are the transverse position coordinate of the beam with wave-vector k on the lens plane and f is the focal length of the lens. This approximation holds under the condition of $x,y<<f_{a}$. However, as evident from Fig. \ref{Figure 1}(b), the thick lens of small focal length used to collect higher order diffracted beams of the object of smaller gratings (less than 10 $\mu m$ or so) needs modification to the phase transformation factor accounting non-paraxial condition. Without any approximation, the phase transfer function of the thick plano-convex lens with a radius of curvature $R$ can be written as,
\begin{equation}\label{non-paraxial}
U(x, y) \sim \exp \left[\frac{i k}{f_{a}} R^{2}\left[1-\sqrt{1-\frac{x^{2}+y^{2}}{R^{2}}}\right]\right]
\end{equation}
It is known that the effective focal length, $f'_{a}$, under the non-paraxial conditions, is smaller than the focal length $f_{a}$ at paraxial conditions. Thus, the multiplicative phase factor of the thick lens under non-paraxial conditions can be written as,
\begin{equation}\label{paraxial}
U(x, y)\sim \exp \left[\frac{i k}{2 f'_{a}}\left(x^{2}+y^{2}\right)\right].
\end{equation}
Comparing Eqs. \ref{non-paraxial} and \ref{paraxial}, we can write the modified focal length of the thick lens from the transition from paraxial to non-paraxial and associated spherical aberration as,
\begin{equation}\label{focal length_theory}
f'_{a} = \frac{f_{a}\left(x^{2}+y^{2}\right)}{2R^{2}[1-\sqrt{1-\frac{x^{2}+y^{2}}{R^{2}}}]}.
\end{equation}
In the presence of 1D periodic objects, all the above equations can be transformed to 1D dimension by making one coordinate zero (here, we consider $y$=0). Again, using the diffraction equation of the grating of the period, $\Lambda$, and illuminating wavelength, $\lambda$, we can find the value of $x$  as,
\begin{equation}\label{diffraction}
    \frac{x}{f_{a}} = \tan(\sin^{-1}({\frac{\lambda}{\Lambda}}))
\end{equation}
Using the value of $x$ in Eq. \ref{focal length_theory} and $y$ = 0, we can numerically calculate the effective focal length of the thick lens as a function of period, $\Lambda$, of the 1D grating. On the other hand, the Talbot length of the 1D grating for the schematic of Fig. \ref{Figure 1}(b), can be repeated as 
\begin{equation}\label{Talbot_length_abberation}
Z'_{T} = \left(\frac{f_b}{f'_a}\right)^2 \left(\frac{2\Lambda^2}{\lambda}\right)
\end{equation}
due to the magnification of the 1D grating at the image plane by a factor, $(\frac{f_{b}}{f'_{a}})$, the ratio of focal length of lens, $L_b$ and effective focal length, $f'_{a}$, of lens, $L_a$, under non-paraxial condition. 
Using Eq. \ref{Talbot_length_abberation}, we can represent the effective focal length of lens $L_a$ as, 
\begin{equation}\label{focal length Experimental}
f'_{a} = \frac{f_{b}\Lambda}{\sqrt{\frac{\lambda Z'_{T}}{2}}}.
\end{equation}

It is evident from Eq. \ref{focal length Experimental} that using a test target 1D grating of known periods and measuring the corresponding Talbot length, we can find the change in the focal length of a thick lens under non-paraxial conditions or the spherical aberration.
\begin{figure}[ht]
\centering
\includegraphics[width=\linewidth]{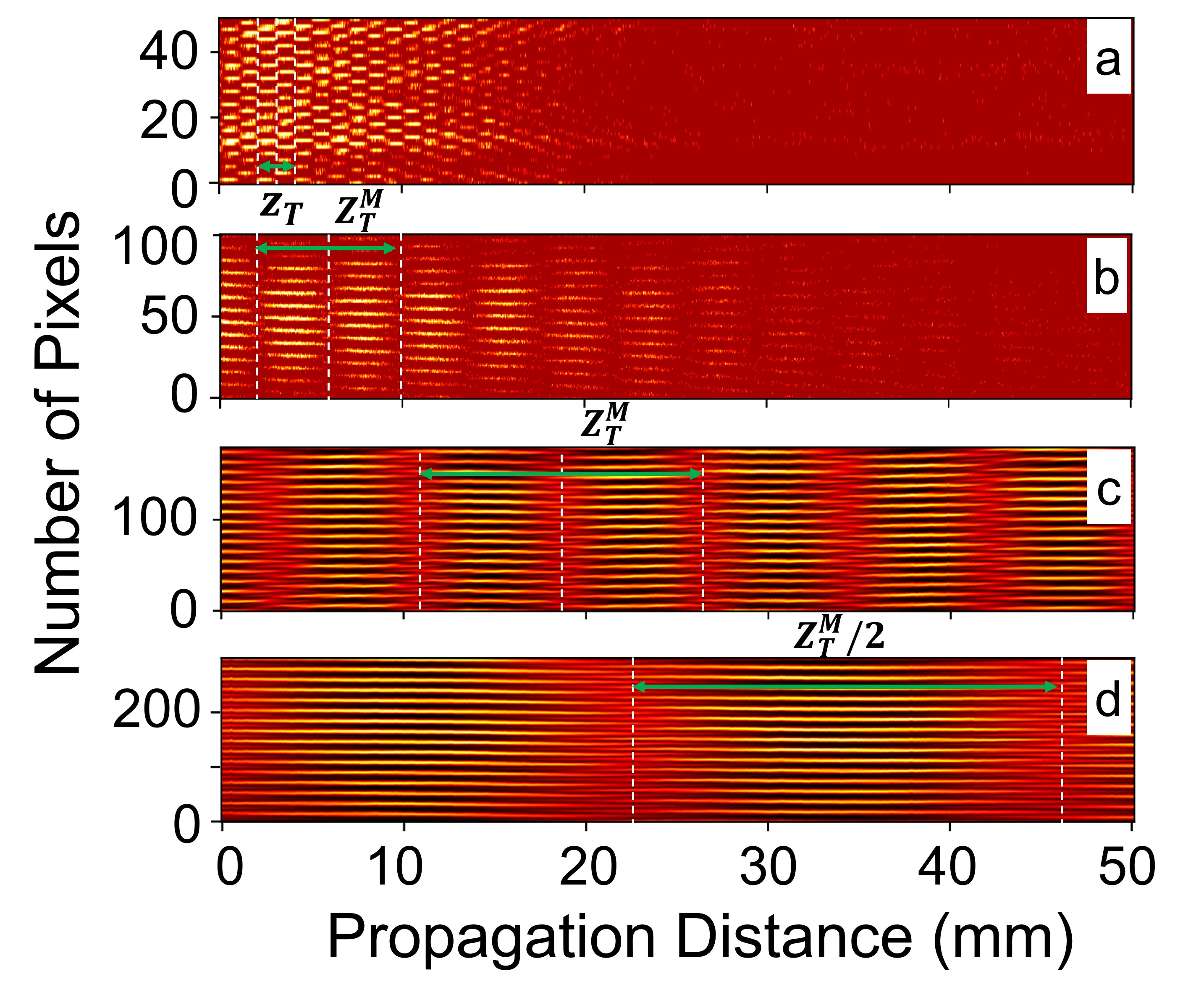}
\caption{Observation of light carpet and subsequent Talbot length change for different positions of the periodic object, (a) 4f imaging, (b) $d$ = 100 mm, (c) $d$ = 70 mm, and (d) $d$ = 40 mm.}
\label{tunable}
\end{figure}

\begin{figure}
\centering
\includegraphics[width=\linewidth]{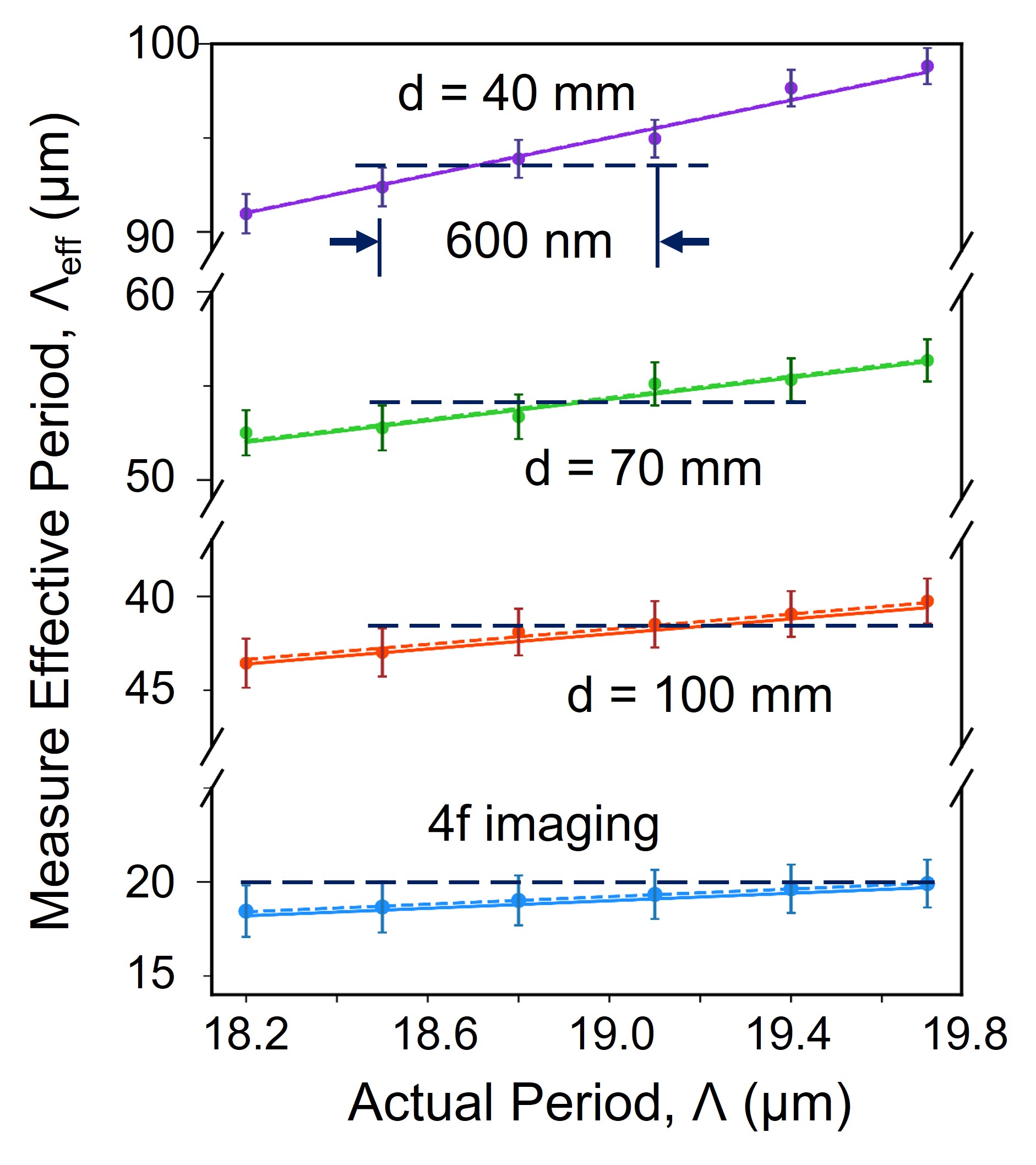}
\caption{Variation of effective periodicity, $\Lambda_{eff}$  as a function of the actual periodicity, $\Lambda$ of the multi-grating crystal, for different position of the object satisfying 4$f$ imaging condition, $d$ = 100 mm, 70 mm and 40 mm. The solid lines are the theoretical results derived using the experimental parameters in Eq.~\ref{Eq. 3}. The black dash line and black arrows are used to identify the measurement resolution.}
\label{periodicity graph}
\end{figure}
We first characterized the experimental setup shown in Fig. \ref{Figure 1}(a) to control the Talbot length by varying the object position, $d$, as described by Eq. \ref{Eq. 4}, particularly for gratings with a lower period than previously reported \cite{Harshit24}. While 
$d$ can be adjusted continuously, we selected three discrete values to verify the tunable Talbot length for a fixed grating period and laser wavelength. Using a He-Ne laser at 632.8 nm and lens, $L_a$ and $L_b$ of focal length $f_a = f_b$ = 200 mm, we placed a 1D periodic object, one-dimensional Ronchi grating (R1S1L1N Negative test target, Thorlabs) of period $25$ $\mu m$ at different positions satisfying the 4$f$ imaging condition, $d$ = 100 mm, 70 mm, and 40 mm. We recorded the intensity profiles using a CCD camera along the propagation direction with a distance step of 0.01, 0.05, 0.05, and 0.1 mm for 4$f$-imaging condition, $d$ = 100 mm, 70 mm, and 40 mm, respectively. The light carpet was then constructed by concatenating the intensity line profiles from the CCD images along a 50 mm propagation distance from a fixed pixel line with a pixel width of 6.45 $\mu$m, consistent with Ref. \cite{Harshit24}.

The results are presented in Fig. \ref{tunable}. The theoretical Talbot length for a period of $\Lambda$ = 25 $\mu m$, and the laser wavelength of $\lambda$ = 632.8 nm is $z_{T}$ = 2$\Lambda$$^{2}$/$\lambda$ =1.977 mm.
According to Eq. \ref{Eq. 4}, the modified Talbot length for unit magnification remains $z_{T}$. As shown in Fig. \ref{tunable}(a), the experimentally measured light carpet for the 25 $\mu$m grating period spans over a propagation distance of approximately 20 mm and clearly identifies multiple Talbot planes. The measured Talbot length was 1.97 mm $\pm$ 0.08 mm, closely matching the theoretical value. For $d$ = 100 mm (Fig. \ref{tunable}(b)), the light carpet extends over a longer propagation distance of 40 mm, with multiple Talbot planes and a magnified Talbot length of 7.93 $\pm$ 0.15 mm. Similarly, as seen in Figs. \ref{tunable}(c) and (d) for $d$ = 70 mm and $d$ = 40 mm, the light carpet extends over the entire 50 mm propagation distance, with three and one Talbot planes observed, corresponding to Talbot lengths of 15.84 $\pm$ 0.2 mm and 49 $\pm$ 0.3 mm, respectively. The Talbot length magnification factors were 4.02$\times$, 8.04$\times$, and 24.87$\times$, resulting from $f_b/d$ = 2.00, 2.83, and 4.99 for $d$ = 100 mm, 70 mm, and 40 mm, respectively. This enhancement in Talbot length could significantly improve the measurement resolution of Talbot-based sensors, enabling the detection of small changes in the period or other physical parameters that influence the period of a periodic object.

According to Eq. \ref{Eq. 4}, we can, in principle, further reduce the $d$ value below 40 mm and produce longer Talbot lengths. However, in a separate experiment and numerical simulation, we observed the Talbot effect for the illumination spot size as small as 5$\times$ the grating period. In such cases, only one Talbot length was observed, followed by strong beam divergence. Therefore, we limited our study to an illumination spot size of more than 15$\times$$\Lambda$, achieved with 
$d$ = 40 mm. An alternative approach to further increase the Talbot length could involve selecting appropriate focal lengths for lenses 
$L_a$ and $L_b$ to achieve a higher $f_b/f_a$ ratio. However, this would result in a fixed Talbot length magnification for a given set of lenses, potentially undermining the goal of achieving a controlled Talbot length, which is the central focus of this study.

\begin{table*}
\caption{\label{table_result}Measured grating periods for different positions of the object}
\begin{ruledtabular}
\begin{tabular}{ccccc}
\multirow{2}{*}{Acutal Grating period, $\Lambda$ ($\mu$m)} & \multicolumn{4}{c}{Measured grating period from experiment, $ \frac{d}{f_b} \Lambda_{\text{eff}}$ ($\mu$m)} \\
\cmidrule(lr){2-5}
 & $d = 40$ mm & $d = 70$ mm & $d = 100$ mm & 4f imaging \\
\midrule
\hline
18.2 & 18.19 ± 0.21 & 18.34 ± 0.42 & 18.10 ± 0.65 & 18.45 ± 1.37 \\

18.5 & 18.47 ± 0.20 & 18.44 ± 0.42 & 18.39 ± 0.64 & 18.67 ± 1.35 \\
18.8 & 18.77 ± 0.20 & 18.64 ± 0.41 & 18.92 ± 0.62 & 19.02 ± 1.33 \\
19.1 & 18.98 ± 0.20 & 19.26 ± 0.40 & 19.13 ± 0.61 & 19.34 ± 1.31 \\
19.4 & 19.52 ± 0.19 & 19.33 ± 0.40 & 19.40 ± 0.60 & 19.64 ± 1.29 \\
19.7 & 19.76 ± 0.19 & 19.69 ± 0.39 & 19.74 ± 0.59 & 19.91 ± 1.27 \\
\end{tabular}
\end{ruledtabular}
\end{table*}

By establishing reliable control over the Talbot length of a periodic object with a fixed grating period (as small as 25 $\mu$m and a constant illumination wavelength, we demonstrate the capability of a Talbot-based sensor to measure small changes in the grating period. In the absence of a suitable periodic object, we used a 1D multi-grating periodically poled Lithium Niobate (PPLN) crystal with the grating period varying from 18.2 $\mu$m to 19.7 $\mu$m with an increment of 300 nm. Placing the crystal at different positions satisfying the 4$f$ imaging condition, $d$ = 100 mm, 70 mm, and 40 mm, we varied the grating period by moving the crystal in the transverse plane and measured the Talbot length from the intensity profiles recorded using the CCD camera along the propagation. Experimentally, for the grating position of 4$f$ condition, $d$ = 100 mm, 70 mm, and 40 mm, we measured respective Talbot length variation of 1.07 - 1.25, 4.20 - 4.99, 8.71 - 10.04 and 26.15 - 30.87 mm for the grating period variation from 18.2 - 19.7 $\mu$m. As evident, we observe a large Talbot length variation (4.72 mm) for a grating period variation of 1.5 $\mu$m while placing the object at $d$ = 40 mm, indicating the possibility of high measurement resolution for a Talbot-based sensor. Using Eq. \ref{Eq. 4} and the measured Talbot length $Z_{T}^{M}$, we have calculated the effective grating period ($\Lambda_{eff}$) for the change of period $\Lambda$ with the results shown in Fig. \ref{periodicity graph}.

As shown in Fig. \ref{periodicity graph}, the experimentally measured effective Talbot length, $\Lambda_{eff}$, varies linearly with the actual grating period, $\Lambda$, over ranges of 18.44 - 19.91, 36.44 - 39.75, 52.50 - 56.36, and 90.97 - 98.82 $\mu$m. Using Eq. \ref{Eq. 4}, the fitting (solid line) to the experimental data (dots) demonstrates a close agreement between theory and experiment, with slopes of 1, 2, 2.85, and 5, precisely matching the values of ($f_b/d$) for changes in the grating period from 18.2-19.7 $\mu$m, and the object positions satisfying the 4$f$ imaging condition, $d$ = 100 mm, 70 mm, and 40 mm. It is also interesting to note that the measured effective grating period lies within the measurement error bar for all $d$ values except $d$ = 40 mm. In fact, for $d$ = 40 mm, as evident from the purple line and dots of Fig. \ref{periodicity graph}, one can clearly distinguish two grating periods (see the black dash and black arrows) with a difference of 600 nm.
However, to get a better perspective on the resolution of the Talbot-based sensor, we have summarized the measurement data in Table \ref{table_result}. As evident from the table~\ref{table_result}, the measured grating period for the object position satisfying 4$f$-imaging condition has an error of $\pm$ 1.3 $\mu$m. Such high error can be attributed to the measurement inaccuracy resulting from the lower resolution ($\sim$10 µm) of the transnational stage used to measure the relatively small Talbot length of the grating under the 4$f$-imaging condition. On the other hand, the increase in Talbot length due to the use of shorter $d$ values — 100 mm, 70 mm, and 40 mm — we observe an improvement in measurement accuracy from $\pm$0.65 $\mu$m, $\pm$0.42 $\mu$m to $\pm$0.21 $\mu$m, respectively. 

In fact, for $d$ = 40 mm, the Talbot-based sensor can measure a change in the grating period as small as $\sim$450 nm for a grating period of 18.2 $\mu$m. It is also evident from the Table that for a constant value $d$, the measurement error or the resolution, as also evident from Eq. \ref{Eq. 5}, improves with the increase in the grating period. For example, as evident from the second column of Table \ref{table_result}, the measurement error improves from $\pm$0.21 $\mu$m to $\pm$ 0.19 $\mu$m for the increase of grating period from 18.2 $\mu$m to 19.7 $\mu$m. These results establish the reliability of the Talbot-based sensor in measuring changes in the grating period, or any physical parameter influencing the grating period of an object, by 450 nm without requiring complex microscopic measurement systems. While large magnification factors in microscopic systems involving intricate lens designs and aberration control have traditionally been necessary for such precision, the current results demonstrate the potential for achieving similar measurements for a periodic object systematically placed between two long focal-length lenses.

\begin{figure}[h]
\centering
\includegraphics[width=\linewidth]{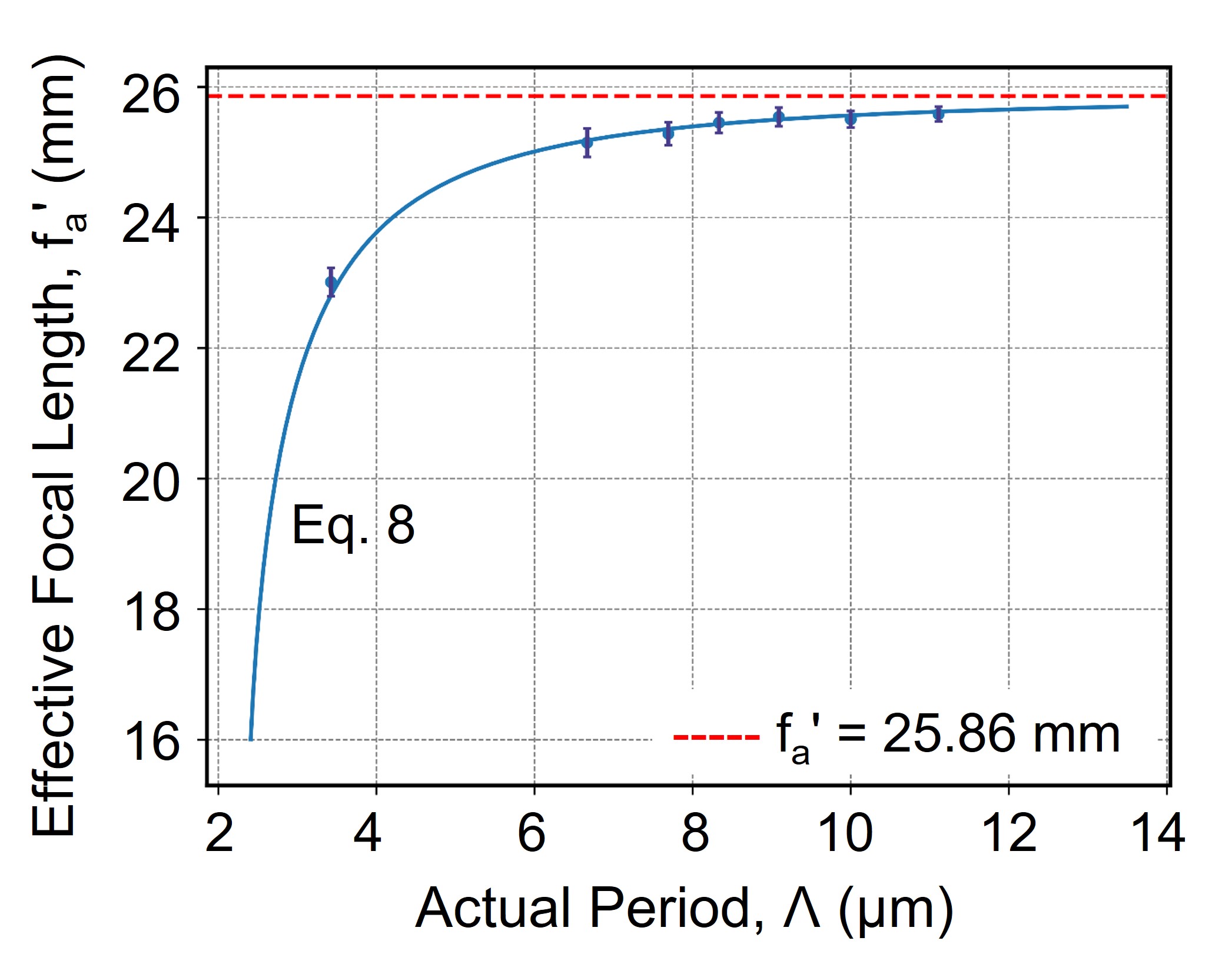}
\caption{Variation of the effective focal length of the thick lens under non-paraxial illumination. The dotted line represents the actual focal length of the lens specified by the vendor. The solid line represents the numerically calculated effective focal length using experimental parameters in Eq. \ref{focal length_theory} and \ref{diffraction}}
\label{aberration_experimental}
\end{figure} 

We further use the Talbot effect to measure the effective focal length of a thick lens due to non-paraxial illumination or spherical aberration. To control the non-paraxial illumination to the lens, $L_a$ we used Ronchi gratings (R1S1L1N negative test target, Thorlabs) as the periodic object with known periods of 11.11 $\mu m$, 10 $\mu m$, 9.09 $\mu m$, 8.33 $\mu m$, 7.69 $\mu m$ and 6.67 $\mu m$ respectively. Due to the unavailability of suitable test grating with a lower period, we have used a periodically poled KTP (PPKTP) crystal of a known grating period of 3.425 $\mu$m. The experimental scheme is shown in Fig. \ref{Figure 1}(b), where the periodic object is imaged using lenses, $L_a$ and $L_b$ in 2$f_a$-2$f_b$ combination. The thick lens, $L_a$, has a focal length, $f_{a}$ = 25.86 mm (for 1064 nm wavelength, from Thorlabs data sheet) and an aperture size of 25.4 mm. We used lenses, $L_{b}$, of two different focal lengths, i.e., 304.95 mm (at 1064 nm, Thorlabs) for grating periods 11.11 $\mu$m to 6.67 $\mu$m and 507.86 mm (at 1064 nm, Thorlabs) for 3.425 $\mu$m grating, for appropriate magnification. Using the laser wavelength of 1064 nm, we recorded the light carpet of the object of different grating periods from the image plane of the lens, $L_b$, and calculated the Talbot lengths. Using the experimentally measured Talbot length for different grating periods, we have calculated the change in the effective focal length of the lens, $L_a$, using Eq. \ref{focal length Experimental}. The results are shown in Fig. \ref{aberration_experimental}. As evident from Fig. \ref{aberration_experimental}, the effective focal length of the thick lens, $L_a$, is lower than the vendor-specified focal length of $f_a$ = 25.86 mm (red dotted line) at paraxial illumination. In fact, with the decrease of the grating period from 11.11 $\mu$m to 3.425 $\mu$m  resulting in the increase of diffraction angle and non-paraxial illumination to the thick lens, the effective focal length, $f'_a$, decreases from 25.58 mm to 23.01 mm. Using the experimental parameters in Eq. \ref{focal length_theory} and Eq. \ref{diffraction}, we numerically calculate the value of effective focal length, $f'_a$ as a function of the grating period, $\Lambda$ with the result shown by the solid line of Fig. \ref{focal length Experimental} in close agreement with the experimental data (dots). From this study, it is evident that the Talbot effect can be effectively used to accurately measure changes in the focal length of a thick lens caused by spherical aberration under non-paraxial illumination. Additionally, the present study establishes the need for accurate focal lengths of the imaging systems used for the study of the non-paraxial Talbot effect. 


In conclusion, we have developed a versatile experimental scheme for Talbot effect-based sensors using a suitable theoretical framework, enabling continuous control of Talbot lengths for periodic structures with micrometer-scale periods. By employing a grating with a period of 25 µm, we achieved tunable Talbot lengths ranging from 1.97 mm to 49 mm, an enhancement of up to 25 times compared to the original range. This tunable Talbot length allowed the sensor to detect periodicity variations of 450 nm for a grating period as small as 18.2 µm with a 632.8 nm wavelength laser, demonstrating the capability to operate in the subwavelength regime without relying on microscopic systems. Furthermore, we derived and experimentally validated a mathematical expression for the effective focal length of a thick lens, accounting for spherical aberration in non-paraxial Talbot imaging. This advancement opens new possibilities for employing Talbot-based sensors to measure various physical parameters that influence the period of periodic objects in real-life applications.

\section*{ACKNOWLEDGMENTS}
Both the authors acknowledge the support of the Dept. of Space, Govt. of India.

\section*{AUTHOR DECLARATIONS}
\subsection*{Conflict of Interest}
The authors have no conflicts to disclose.
\subsection*{Data Availability Statement}
The data that support the findings of this study are available from the corresponding author upon reasonable request.

\bibliography{reference}

\begin{thebibliography}{25}
\expandafter\ifx\csname natexlab\endcsname\relax\def\natexlab#1{#1}\fi
\expandafter\ifx\csname bibnamefont\endcsname\relax
  \def\bibnamefont#1{#1}\fi
\expandafter\ifx\csname bibfnamefont\endcsname\relax
  \def\bibfnamefont#1{#1}\fi
\expandafter\ifx\csname citenamefont\endcsname\relax
  \def\citenamefont#1{#1}\fi
\expandafter\ifx\csname url\endcsname\relax
  \def\url#1{\texttt{#1}}\fi
\expandafter\ifx\csname urlprefix\endcsname\relax\def\urlprefix{URL }\fi
\providecommand{\bibinfo}[2]{#2}
\providecommand{\eprint}[2][]{\url{#2}}

\bibitem[{\citenamefont{Talbot}(1836)}]{talbot1836lxxvi}
\bibinfo{author}{\bibfnamefont{H.~F.} \bibnamefont{Talbot}}, \bibinfo{journal}{The London, Edinburgh, and Dublin Philosophical Magazine and Journal of Science} \textbf{\bibinfo{volume}{9}}, \bibinfo{pages}{401} (\bibinfo{year}{1836}).

\bibitem[{\citenamefont{Rayleigh}(1881)}]{rayleigh1881xxv}
\bibinfo{author}{\bibfnamefont{L.}~\bibnamefont{Rayleigh}}, \bibinfo{journal}{The London, Edinburgh, and Dublin Philosophical Magazine and Journal of Science} \textbf{\bibinfo{volume}{11}}, \bibinfo{pages}{196} (\bibinfo{year}{1881}).

\bibitem[{\citenamefont{Berry and Klein}(1996)}]{Berry:96}
\bibinfo{author}{\bibfnamefont{M.~V.} \bibnamefont{Berry}} \bibnamefont{and} \bibinfo{author}{\bibfnamefont{S.}~\bibnamefont{Klein}}, \bibinfo{journal}{Journal of Modern Optics} \textbf{\bibinfo{volume}{43}}, \bibinfo{pages}{2139} (\bibinfo{year}{1996}).

\bibitem[{\citenamefont{Wen et~al.}(2013)\citenamefont{Wen, Zhang, and Xiao}}]{wen2013talbot}
\bibinfo{author}{\bibfnamefont{J.}~\bibnamefont{Wen}}, \bibinfo{author}{\bibfnamefont{Y.}~\bibnamefont{Zhang}}, \bibnamefont{and} \bibinfo{author}{\bibfnamefont{M.}~\bibnamefont{Xiao}}, \bibinfo{journal}{Advances in Optics and Photonics} \textbf{\bibinfo{volume}{5}}, \bibinfo{pages}{83} (\bibinfo{year}{2013}).

\bibitem[{\citenamefont{Kung et~al.}(2001)\citenamefont{Kung, Bhatnagar, and Miller}}]{kung2001transform}
\bibinfo{author}{\bibfnamefont{H.~L.} \bibnamefont{Kung}}, \bibinfo{author}{\bibfnamefont{A.}~\bibnamefont{Bhatnagar}}, \bibnamefont{and} \bibinfo{author}{\bibfnamefont{D.~A.} \bibnamefont{Miller}}, \bibinfo{journal}{Optics Letters} \textbf{\bibinfo{volume}{26}}, \bibinfo{pages}{1645} (\bibinfo{year}{2001}).

\bibitem[{\citenamefont{Ye et~al.}(2016)\citenamefont{Ye, Atabaki, Han, and Ram}}]{ye2016miniature}
\bibinfo{author}{\bibfnamefont{E.}~\bibnamefont{Ye}}, \bibinfo{author}{\bibfnamefont{A.~H.} \bibnamefont{Atabaki}}, \bibinfo{author}{\bibfnamefont{N.}~\bibnamefont{Han}}, \bibnamefont{and} \bibinfo{author}{\bibfnamefont{R.~J.} \bibnamefont{Ram}}, \bibinfo{journal}{Optics Letters} \textbf{\bibinfo{volume}{41}}, \bibinfo{pages}{2434} (\bibinfo{year}{2016}).

\bibitem[{\citenamefont{Podanchuk et~al.}(2016{\natexlab{a}})\citenamefont{Podanchuk, Goloborodko, Kotov, Kovalenko, Kurashov, and Dan'ko}}]{Podanchuk:16}
\bibinfo{author}{\bibfnamefont{D.~V.} \bibnamefont{Podanchuk}}, \bibinfo{author}{\bibfnamefont{A.~A.} \bibnamefont{Goloborodko}}, \bibinfo{author}{\bibfnamefont{M.~M.} \bibnamefont{Kotov}}, \bibinfo{author}{\bibfnamefont{A.~V.} \bibnamefont{Kovalenko}}, \bibinfo{author}{\bibfnamefont{V.~N.} \bibnamefont{Kurashov}}, \bibnamefont{and} \bibinfo{author}{\bibfnamefont{V.~P.} \bibnamefont{Dan'ko}}, \bibinfo{journal}{Applied Optics} \textbf{\bibinfo{volume}{55}}, \bibinfo{pages}{B150} (\bibinfo{year}{2016}{\natexlab{a}}).

\bibitem[{\citenamefont{Wang et~al.}(2009)\citenamefont{Wang, Gill, and Molnar}}]{Wang:09}
\bibinfo{author}{\bibfnamefont{A.}~\bibnamefont{Wang}}, \bibinfo{author}{\bibfnamefont{P.}~\bibnamefont{Gill}}, \bibnamefont{and} \bibinfo{author}{\bibfnamefont{A.}~\bibnamefont{Molnar}}, \bibinfo{journal}{Applied Optics} \textbf{\bibinfo{volume}{48}}, \bibinfo{pages}{5897} (\bibinfo{year}{2009}).

\bibitem[{\citenamefont{Spagnolo and Ambrosini}(2000)}]{spagnolo2000talbot}
\bibinfo{author}{\bibfnamefont{G.~S.} \bibnamefont{Spagnolo}} \bibnamefont{and} \bibinfo{author}{\bibfnamefont{D.}~\bibnamefont{Ambrosini}}, \bibinfo{journal}{Measurement Science and Technology} \textbf{\bibinfo{volume}{11}}, \bibinfo{pages}{77} (\bibinfo{year}{2000}).

\bibitem[{\citenamefont{Cao et~al.}(2022)\citenamefont{Cao, Zhang, Niu, Ma, Yang, Li, and Xin}}]{Cao:22}
\bibinfo{author}{\bibfnamefont{B.}~\bibnamefont{Cao}}, \bibinfo{author}{\bibfnamefont{R.}~\bibnamefont{Zhang}}, \bibinfo{author}{\bibfnamefont{Q.}~\bibnamefont{Niu}}, \bibinfo{author}{\bibfnamefont{X.}~\bibnamefont{Ma}}, \bibinfo{author}{\bibfnamefont{Z.}~\bibnamefont{Yang}}, \bibinfo{author}{\bibfnamefont{M.}~\bibnamefont{Li}}, \bibnamefont{and} \bibinfo{author}{\bibfnamefont{C.}~\bibnamefont{Xin}}, \bibinfo{journal}{Applied Optics} \textbf{\bibinfo{volume}{61}}, \bibinfo{pages}{9873} (\bibinfo{year}{2022}).

\bibitem[{\citenamefont{Testorf et~al.}(1996)\citenamefont{Testorf, Jahns, Khilo, and Goncharenko}}]{TESTORF1996167}
\bibinfo{author}{\bibfnamefont{M.}~\bibnamefont{Testorf}}, \bibinfo{author}{\bibfnamefont{J.}~\bibnamefont{Jahns}}, \bibinfo{author}{\bibfnamefont{N.~A.} \bibnamefont{Khilo}}, \bibnamefont{and} \bibinfo{author}{\bibfnamefont{A.~M.} \bibnamefont{Goncharenko}}, \bibinfo{journal}{Optics Communications} \textbf{\bibinfo{volume}{129}}, \bibinfo{pages}{167} (\bibinfo{year}{1996}).

\bibitem[{\citenamefont{Sekine et~al.}(2006)\citenamefont{Sekine, Shibuya, Ukai, Komatsu, Hattori, Mihashi, Nakazawa, and Hirohara}}]{sekine2006measurement}
\bibinfo{author}{\bibfnamefont{R.}~\bibnamefont{Sekine}}, \bibinfo{author}{\bibfnamefont{T.}~\bibnamefont{Shibuya}}, \bibinfo{author}{\bibfnamefont{K.}~\bibnamefont{Ukai}}, \bibinfo{author}{\bibfnamefont{S.}~\bibnamefont{Komatsu}}, \bibinfo{author}{\bibfnamefont{M.}~\bibnamefont{Hattori}}, \bibinfo{author}{\bibfnamefont{T.}~\bibnamefont{Mihashi}}, \bibinfo{author}{\bibfnamefont{N.}~\bibnamefont{Nakazawa}}, \bibnamefont{and} \bibinfo{author}{\bibfnamefont{Y.}~\bibnamefont{Hirohara}}, \bibinfo{journal}{Optical Review} \textbf{\bibinfo{volume}{13}}, \bibinfo{pages}{207} (\bibinfo{year}{2006}).

\bibitem[{\citenamefont{Podanchuk et~al.}(2016{\natexlab{b}})\citenamefont{Podanchuk, Goloborodko, Kotov, Kovalenko, Kurashov, and Dan’ko}}]{podanchuk2016adaptive}
\bibinfo{author}{\bibfnamefont{D.~V.} \bibnamefont{Podanchuk}}, \bibinfo{author}{\bibfnamefont{A.~A.} \bibnamefont{Goloborodko}}, \bibinfo{author}{\bibfnamefont{M.~M.} \bibnamefont{Kotov}}, \bibinfo{author}{\bibfnamefont{A.~V.} \bibnamefont{Kovalenko}}, \bibinfo{author}{\bibfnamefont{V.~N.} \bibnamefont{Kurashov}}, \bibnamefont{and} \bibinfo{author}{\bibfnamefont{V.~P.} \bibnamefont{Dan’ko}}, \bibinfo{journal}{Applied Optics} \textbf{\bibinfo{volume}{55}}, \bibinfo{pages}{B150} (\bibinfo{year}{2016}{\natexlab{b}}).

\bibitem[{\citenamefont{Ring et~al.}(2012)\citenamefont{Ring, Lindberg, Howls, and Dennis}}]{ring2012aberration}
\bibinfo{author}{\bibfnamefont{J.~D.} \bibnamefont{Ring}}, \bibinfo{author}{\bibfnamefont{J.}~\bibnamefont{Lindberg}}, \bibinfo{author}{\bibfnamefont{C.~J.} \bibnamefont{Howls}}, \bibnamefont{and} \bibinfo{author}{\bibfnamefont{M.~R.} \bibnamefont{Dennis}}, \bibinfo{journal}{Journal of Optics} \textbf{\bibinfo{volume}{14}}, \bibinfo{pages}{075702} (\bibinfo{year}{2012}).

\bibitem[{\citenamefont{Sunday et~al.}(2015)\citenamefont{Sunday, List, Chawla, and Kline}}]{sunday2015determining}
\bibinfo{author}{\bibfnamefont{D.~F.} \bibnamefont{Sunday}}, \bibinfo{author}{\bibfnamefont{S.}~\bibnamefont{List}}, \bibinfo{author}{\bibfnamefont{J.~S.} \bibnamefont{Chawla}}, \bibnamefont{and} \bibinfo{author}{\bibfnamefont{R.~J.} \bibnamefont{Kline}}, \bibinfo{journal}{Journal of Applied Crystallography} \textbf{\bibinfo{volume}{48}}, \bibinfo{pages}{1355} (\bibinfo{year}{2015}).

\bibitem[{\citenamefont{Deng et~al.}(2020)\citenamefont{Deng, Shapira, Remez, Li, and Arie}}]{Deng:20}
\bibinfo{author}{\bibfnamefont{Z.}~\bibnamefont{Deng}}, \bibinfo{author}{\bibfnamefont{N.}~\bibnamefont{Shapira}}, \bibinfo{author}{\bibfnamefont{R.}~\bibnamefont{Remez}}, \bibinfo{author}{\bibfnamefont{Y.}~\bibnamefont{Li}}, \bibnamefont{and} \bibinfo{author}{\bibfnamefont{A.}~\bibnamefont{Arie}}, \bibinfo{journal}{Optics Letters} \textbf{\bibinfo{volume}{45}}, \bibinfo{pages}{2538} (\bibinfo{year}{2020}).

\bibitem[{\citenamefont{Rogers et~al.}(2012)\citenamefont{Rogers, Lindberg, Roy, Savo, Chad, Dennis, and Zheludev}}]{rogers2012super}
\bibinfo{author}{\bibfnamefont{E.~T.} \bibnamefont{Rogers}}, \bibinfo{author}{\bibfnamefont{J.}~\bibnamefont{Lindberg}}, \bibinfo{author}{\bibfnamefont{T.}~\bibnamefont{Roy}}, \bibinfo{author}{\bibfnamefont{S.}~\bibnamefont{Savo}}, \bibinfo{author}{\bibfnamefont{J.~E.} \bibnamefont{Chad}}, \bibinfo{author}{\bibfnamefont{M.~R.} \bibnamefont{Dennis}}, \bibnamefont{and} \bibinfo{author}{\bibfnamefont{N.~I.} \bibnamefont{Zheludev}}, \bibinfo{journal}{Nature Materials} \textbf{\bibinfo{volume}{11}}, \bibinfo{pages}{432} (\bibinfo{year}{2012}).

\bibitem[{\citenamefont{Huang et~al.}(2007)\citenamefont{Huang, Zheludev, Chen, and Javier Garcia~de Abajo}}]{huang2007focusing}
\bibinfo{author}{\bibfnamefont{F.~M.} \bibnamefont{Huang}}, \bibinfo{author}{\bibfnamefont{N.}~\bibnamefont{Zheludev}}, \bibinfo{author}{\bibfnamefont{Y.}~\bibnamefont{Chen}}, \bibnamefont{and} \bibinfo{author}{\bibfnamefont{F.}~\bibnamefont{Javier Garcia~de Abajo}}, \bibinfo{journal}{Applied Physics Letters} \textbf{\bibinfo{volume}{90}} (\bibinfo{year}{2007}).

\bibitem[{\citenamefont{Photia et~al.}(2019)\citenamefont{Photia, Temnuch, Srisuphaphon, Tanasanchai, Anukool, Wongrach, Manit, Chiangga, and Deachapunya}}]{Photia:19}
\bibinfo{author}{\bibfnamefont{T.}~\bibnamefont{Photia}}, \bibinfo{author}{\bibfnamefont{W.}~\bibnamefont{Temnuch}}, \bibinfo{author}{\bibfnamefont{S.}~\bibnamefont{Srisuphaphon}}, \bibinfo{author}{\bibfnamefont{N.}~\bibnamefont{Tanasanchai}}, \bibinfo{author}{\bibfnamefont{W.}~\bibnamefont{Anukool}}, \bibinfo{author}{\bibfnamefont{K.}~\bibnamefont{Wongrach}}, \bibinfo{author}{\bibfnamefont{P.}~\bibnamefont{Manit}}, \bibinfo{author}{\bibfnamefont{S.}~\bibnamefont{Chiangga}}, \bibnamefont{and} \bibinfo{author}{\bibfnamefont{S.}~\bibnamefont{Deachapunya}}, \bibinfo{journal}{Applied Optics} \textbf{\bibinfo{volume}{58}}, \bibinfo{pages}{270} (\bibinfo{year}{2019}).

\bibitem[{\citenamefont{Arrieta et~al.}(2019)\citenamefont{Arrieta, Bolognini, and Torres}}]{ARRIETA2019988}
\bibinfo{author}{\bibfnamefont{E.}~\bibnamefont{Arrieta}}, \bibinfo{author}{\bibfnamefont{N.}~\bibnamefont{Bolognini}}, \bibnamefont{and} \bibinfo{author}{\bibfnamefont{C.}~\bibnamefont{Torres}}, \bibinfo{journal}{Optik} \textbf{\bibinfo{volume}{183}}, \bibinfo{pages}{988} (\bibinfo{year}{2019}).

\bibitem[{\citenamefont{Hua et~al.}(2012)\citenamefont{Hua, Suh, Zhou, Huntington, and Odom}}]{hua2012talbot}
\bibinfo{author}{\bibfnamefont{Y.}~\bibnamefont{Hua}}, \bibinfo{author}{\bibfnamefont{J.~Y.} \bibnamefont{Suh}}, \bibinfo{author}{\bibfnamefont{W.}~\bibnamefont{Zhou}}, \bibinfo{author}{\bibfnamefont{M.~D.} \bibnamefont{Huntington}}, \bibnamefont{and} \bibinfo{author}{\bibfnamefont{T.~W.} \bibnamefont{Odom}}, \bibinfo{journal}{Optics Express} \textbf{\bibinfo{volume}{20}}, \bibinfo{pages}{14284} (\bibinfo{year}{2012}).

\bibitem[{\citenamefont{Goodman}(2005)}]{goodman2005introduction}
\bibinfo{author}{\bibfnamefont{J.~W.} \bibnamefont{Goodman}}, \emph{\bibinfo{title}{Introduction to Fourier optics}} (\bibinfo{publisher}{Roberts and Company publishers}, \bibinfo{year}{2005}).

\bibitem[{\citenamefont{Bachimanchi et~al.}(2024)\citenamefont{Bachimanchi, Sarkar, Ebrahim-Zadeh, and Samanta}}]{Harshit24}
\bibinfo{author}{\bibfnamefont{H.}~\bibnamefont{Bachimanchi}}, \bibinfo{author}{\bibfnamefont{S.~J.} \bibnamefont{Sarkar}}, \bibinfo{author}{\bibfnamefont{M.}~\bibnamefont{Ebrahim-Zadeh}}, \bibnamefont{and} \bibinfo{author}{\bibfnamefont{G.~K.} \bibnamefont{Samanta}}, \bibinfo{journal}{Optics Express} \textbf{\bibinfo{volume}{32}}, \bibinfo{pages}{15967} (\bibinfo{year}{2024}).

\bibitem[{\citenamefont{Harshith and Samanta}(2019)}]{Harshith2019}
\bibinfo{author}{\bibfnamefont{B.~S.} \bibnamefont{Harshith}} \bibnamefont{and} \bibinfo{author}{\bibfnamefont{G.~K.} \bibnamefont{Samanta}}, \bibinfo{journal}{Scientific Reports} \textbf{\bibinfo{volume}{9}}, \bibinfo{pages}{10916} (\bibinfo{year}{2019}).

\bibitem[{\citenamefont{Jabir et~al.}(2016)\citenamefont{Jabir, Apurv~Chaitanya, Aadhi, and Samanta}}]{Jabir2016}
\bibinfo{author}{\bibfnamefont{M.~V.} \bibnamefont{Jabir}}, \bibinfo{author}{\bibfnamefont{N.}~\bibnamefont{Apurv~Chaitanya}}, \bibinfo{author}{\bibfnamefont{A.}~\bibnamefont{Aadhi}}, \bibnamefont{and} \bibinfo{author}{\bibfnamefont{G.~K.} \bibnamefont{Samanta}}, \bibinfo{journal}{Scientific Reports} \textbf{\bibinfo{volume}{6}}, \bibinfo{pages}{21877} (\bibinfo{year}{2016}).

\end{thebibliography}

\end{document}